\documentclass[conference,final,]{IEEEtran}
\ifCLASSINFOpdf
\else
\fi

\usepackage{graphicx}
\makeatletter
\def\maxwidth{\ifdim\Gin@nat@width>\linewidth\linewidth
\else\Gin@nat@width\fi}
\makeatother
\let\Oldincludegraphics\includegraphics
\renewcommand{\includegraphics}[1]{\Oldincludegraphics[width=\maxwidth]{#1}}

\usepackage[unicode=true]{hyperref}

\hypersetup{
            pdftitle={A Coefficient of Determination for Probabilistic Topic Models},
            pdfborder={0 0 0},
            breaklinks=true}
\usepackage{bm}
\usepackage{amsbsy}
\usepackage{amsmath}
\usepackage{amsfonts}
\providecommand{\tightlist}{%
  \setlength{\itemsep}{0pt}\setlength{\parskip}{0pt}}

\begin{document}
%
\title{A Coefficient of Determination for Probabilistic Topic Models}


\author{


\IEEEauthorblockN{
Tommy Jones 
}
\IEEEauthorblockA{Computational and Data Sciences\\
George Mason University\\
Fairfax, VA
\\jones.thos.w@gmail.com
}


}


%


\maketitle

\begin{abstract}
This research proposes a new (old) metric for evaluating goodness of fit
in topic models, the coefficient of determination, or \(R^2\). Within
the context of topic modeling, \(R^2\) has the same interpretation that
it does when used in a broader class of statistical models. Reporting
\(R^2\) with topic models addresses two current problems in topic
modeling: a lack of standard cross-contextual evaluation metrics for
topic modeling and ease of communication with lay audiences. The author
proposes that \(R^2\) should be reported as a standard metric when
constructing topic models.
\end{abstract}



\maketitle


%
\IEEEpeerreviewmaketitle

\hypertarget{introduction}{%
\section{Introduction}\label{introduction}}

According to an often-quoted but never cited definition, ``the goodness
of fit of a statistical model describes how well it fits a set of
observations. Measures of goodness of fit typically summarize the
discrepancy between observed values and the values expected under the
model in question.''\footnote{This quote appears verbatim on Wikipedia
  and countless books, papers, and websites.} Goodness of fit measures
vary with the goals of those constructing the statistical model.
Inferential goals may emphasize in-sample fit while predictive goals may
emphasize out-of-sample fit. Prior information may be included in the
goodness of fit measure for Bayesian models, or it may not. Goodness of
fit measures may include methods to correct for model overfitting. In
short, goodness of fit measures the performance of a statistical model
against the ground truth of observed data. Fitting the data well is
generally a necessary---though not sufficient---condition for trust in a
statistical model, whatever its goals.

Of course, goodness of fit is only one concern in statistical modeling.
Researchers may trade some goodness of fit for ease of interpretation.
For example, many accurate and robust predictive models are
non-parametric ``black boxes'', making inference {[}human interpretation
of model predictions{]} difficult. Researchers may trade off some
goodness of fit for interpretability by selecting a more restrictive
parametric model. If the emphasis is on prediction, researchers may
trade some in-sample fit for predictive robustness. This is cited as one
motivation for using the Bayesian latent Dirichlet allocation (LDA)
topic model over the frequentist probabilistic latent semantic analysis
(pLSA). It is alleged that pLSA tends to overfit its training sample,
making its estimates fragile to the introduction of new data (Blei, Ng,
and Jordan 2003). Under certain conditions, pLSA and LDA are equivalent
models, however (Girolami and Kabán 2003).

Goodness of fit manifests itself in topic modeling through word
frequencies. It is a common misconception that topic models are
fully-unsupervised methods. If true, this would mean that no
observations exist upon which to compare a model's fitted values.
However, probabilistic topic models are ultimately generative models of
word frequencies (Blei, Ng, and Jordan 2003). The expected value of word
frequencies in a document under a topic model is given by the expected
value of a multinomial random variable. The that can be compared to the
predictions, then, are the word frequencies themselves. Most goodness of
fit measures in topic modeling are restricted to in-sample fit. Yet some
out-of-sample measures have been developed (Buntine 2009).

\hypertarget{probabilistic-topic-models}{%
\section{Probabilistic Topic Models}\label{probabilistic-topic-models}}

Probabilistic topic models are a family of stochastic models for
estimating abstract ``topics'' in a set of documents. Many methods have
been developed to provide a flexible family of topic models. Some
include frequently available metadata about documents, such as the time
of the publication (Blei and Lafferty 2006), or other metadata about the
document or author (Roberts et al. 2014). Most probabilistic topic
models are Bayesian, though probabilistic latent semantic analysis
(pLSA) is frequentist. Without loss of generality, all probabilistic
topic models model the document-generating process as a mixture of
multinomial distributions.\footnote{The terms ``multinomial'' and
  ``categorical'' are often used interchangeably in the topic modeling
  literature. A multinomial distribution is the sum of categorical
  distributions---or multiple trials. To avoid confusion, this paper
  uses multinomial and specifies the number of trials, even if there is
  only one trial.} Probabilistic topic models estimate parameters of an
idealized stochastic process for how words get onto the page. Instead of
writing full, syntactically-coherent, sentences, the author samples a
topic from a multinomial distribution and then, given the topic, samples
a word. The process for a single draw of the \(n\)-th word for the
\(d\)-th document, \(w_{d,n}\), is

\begin{enumerate}
\def\labelenumi{\arabic{enumi}.}
\tightlist
\item
  Sample \(z_{d,n}\sim\) \(\text{Multinomial}(1,\boldsymbol\theta_d)\)
\item
  Sample \(w_{d,n}\sim\)
  \(\text{Multinomial}(1,\boldsymbol\phi_{z_{d,n}})\)
\end{enumerate}

The variable \(z_{d,n}\) is latent. The author repeats this process
\(Nd\) times until the document is ``complete''. For a corpus of \(D\)
documents, \(V\) unique tokens, and \(K\) latent topics, the goal is to
estimate two matrices: \(\boldsymbol\Theta\) and \(\boldsymbol\Phi\).
The \(d\)-th row of \(\boldsymbol\Theta\) comprises
\(\boldsymbol\theta_d\), above. And the \(k\)-th row of
\(\boldsymbol\Phi\) comprises \(\boldsymbol\phi_k\).\footnote{LDA
  estimates these parameters by placing Dirichlet priors on \(\theta_d\)
  and \(\phi_k\).} The document term matrix (DTM)---\(\mathbf{W}\)---can
be thought of as the result of repeated sampling from
\(\boldsymbol\Theta\) and \(\boldsymbol\Phi\). In expectation we have
the following relationship:\footnote{This relationship holds for
  frequentist models. For Bayesian models, where there are priors on
  \(\boldsymbol\Theta\) and \(\boldsymbol\Phi\), the law of total
  expectation applies. This is the method used in the Appendix.}

\begin{align}
  \mathbb{E}(\mathbf{W}) &= \mathbf{n} \odot \Theta \cdot \Phi
\end{align}

Above, \(\mathbf{n}\) is a \(D\)-length vector whose \(d\)-th entry is
the number of terms in the \(d\)-th document and \(\odot\) denotes
elementwise multiplication.

\hypertarget{evaluation-metrics-for-topic-models}{%
\section{Evaluation Metrics for Topic
Models}\label{evaluation-metrics-for-topic-models}}

Most research on topic model evaluation has focused on presenting
ordered lists of words that meet human judgement about words that belong
together. For each topic, words are ordered from the highest value of
\(\boldsymbol\phi_k\) to the lowest (i.e.~the most to least frequent in
each topic). In Chang et al. (2009) the authors introduce the ``intruder
test.'' Judges are shown a few high-probability words in a topic, with
one low-probability word mixed in. Judges must find the low-probability
word, the intruder. They then repeat the procedure with documents
instead of words. A good topic model should allow judges to easily
detect the intruders.

Coherence metrics attempt to approximate the results of intruder tests
in an automated fashion. Researchers have put forward several coherence
measures. These typically compare pairs of highly-ranked words within
topics. Rosner et al. (2014) evaluate several of these. They have human
evaluators rank topics by quality and then compare rankings based on
various coherence measures to the ranking of the evaluators. They
express skepticism that existing coherence measures are sufficient to
assess topic quality. In an ACL paper, Lau, Newman, and Baldwin (2014)
find that normalized pointwise mutual information (NPMI) is a coherence
metric that closely resembles human judgement.

Measuring goodness of fit in topic models has received less attention.
The primary goodness of fit measures in topic modeling are likelihood
metrics. Likelihoods, generally the log likelihood, are naturally
obtained from probabilistic topic models. Likelihoods may contain prior
information, as is often the case with Bayesian models. If prior
information is unknown or undesired, researchers may calculate the
likelihood using only estimated parameters. Researchers have used
likelihoods to select the number of topics (Griffiths and Steyvers
2004), compare priors (Wallach, Mimno, and McCallum 2009), or otherwise
evaluate the efficacy of different modeling procedures. (Asuncion et al.
2009) (Nguyen, Boyd-Graber, and Resnik 2014) A popular likelihood method
for evaluating out-of-sample fit is called perplexity. Perplexity
measures a transformation of the likelihood of the held-out words
conditioned on the trained model. However, Chang et al. (2009),
researchers have eschewed such goodness of fit metrics. This author
believes this is a mistake which---in part---motivates this research
paper.

Researchers have used also used topic models as classifiers. These
evaluations employ precision and recall or the area under a receiver
operator characteristic (ROC) curve (AUC) on topically-tagged corpora
(Asuncion et al. 2009). The most prevalent topic in each document is
taken as a document's topical classification.

Though useful, current evaluation metrics in topic modeling are
difficult to interpret, are inappropriate for use in topic modeling, or
cannot be produced easily. Intruder tests are time-consuming and costly,
making intruder tests infeasible to conduct regularly. Coherence is not
primarily a goodness of fit measure. AUC, precision, and recall metrics
mis-represent topic models as binary classifiers. This misrepresentation
ignores one fundamental motivation for using topic models: allowing
documents to contain multiple topics. This approach also requires
substantial subjective judgement. Researchers must examine the
high-probability words in a topic and decide whether it corresponds to
the corpus topic tags or not.

Likelihoods have an intuitive definition: they represent the probability
of observing the training data if the model is true. Yet properties of
the underlying corpus influence the scale of the likelihood function.
Adding more documents, having a larger vocabulary, and even having
longer documents all reduce the likelihood. Likelihoods of multiple
models on the same corpus can be compared. (Researchers often do this to
help select the number of topics for a final model.)(Griffiths and
Steyvers 2004) Topic models on different corpora cannot be compared,
however. One corpus may have 1,000 documents and 5,000 tokens, while
another may have 10,000 documents and 25,000 tokens. The likelihood of a
model on the latter corpus will be much smaller than a model on the
former. Yet this does not indicate the model on the latter corpus is a
worse fit; the likelihood function is simply on a different scale.
Perplexity is a transformation of the likelihood for out-of-sample
documents. The transformation makes perplexity less intuitive than a raw
likelihood. Perplexity's scale is influenced by the same factors as the
likelihood.

\hypertarget{the-coefficient-of-determination-r2}{%
\section{\texorpdfstring{The Coefficient of Determination:
\(R^2\)}{The Coefficient of Determination: R\^{}2}}\label{the-coefficient-of-determination-r2}}

The coefficient of determination is a popular, intuitive, and
easily-interpretable goodness of fit measure. The coefficient of
determination, denoted \(R^2\), is most common in ordinary least squares
(OLS) regression. However, researchers have developed \(R^2\) and
several pseudo \(R^2\) measures for many classes of statistical models.
The largest value of \(R^2\) is 1, indicating a model fits the data
perfectly. The formal definition of \(R^2\) (below) is
interpreted---without loss of generality---as the proportion of
\emph{variability} in the data that is explained by the model. For
linear models with outcomes in \(\mathbb{R}_1\), \(R^2\) is bound
between 0 and 1 and is the proportion of \emph{variance} in the data
explained by the model (Neter et al. 1996). Even outside of the context
of a linear model, \(R^2\) retains its maximum of 1 and its
interpretation as the proportion of explained variability. Negative
values of \(R^2\) are possible for non-linear models or models in
\(\mathbb{R}_M\) where \(M > 1\). These negative values indicate that
simply guessing the mean outcome is a better fit than the
model.\footnote{In this author's experience, negative values indicate an
  error in one's code, rather than an exceptionally poor fitting model.}

\hypertarget{the-standard-definition-of-r2}{%
\subsection{\texorpdfstring{The Standard Definition of
\(R^2\)}{The Standard Definition of R\^{}2}}\label{the-standard-definition-of-r2}}

For a model, \(f\), of outcome variable, \(y\), where there are \(N\)
observations, \(R^2\) is derived from the following:

\begin{align}
  \bar{y} &= \frac{1}{N}\sum_{i=1}^{N}y_i\\
  SS_{tot.} &= \sum_{i=1}^N{(y_i-\bar{y})^2}\\
  SS_{resid.} &= \sum_{i=1}^N{(f_i-y_i)^2}
\end{align}

Thus, the standard definition of \(R^2\) is a ratio of summed squared
errors.

\begin{align}
    R^2 \equiv 1 - \frac{SS_{resid.}}{SS_{tot.}}
\end{align}

\hypertarget{a-geometric-interpretation-of-r2}{%
\subsection{\texorpdfstring{A Geometric Interpretation of
\(R^2\)}{A Geometric Interpretation of R\^{}2}}\label{a-geometric-interpretation-of-r2}}

\(R^2\) has a geometric interpretation as well. \(SS_{tot.}\) is the
total squared-Euclidean distance from each \(y_i\) to the mean outcome,
\(\bar{y}\). Then \(SS_{resid.}\) is the total squared-Euclidean
distance from each \(y_i\) to its predicted value under the model,
\(f_i\). Recall that for any two points
\(\mathbf{p}, \mathbf{q} \in \mathbb{R}_M\)

\begin{align}
    d(\mathbf{p},\mathbf{q}) = \sqrt{\sum_{i=1}^M{(p_i - q_i)^2}}
\end{align}

where \(d(\mathbf{p}, \mathbf{q})\) denotes the Euclidean distance
between \(\mathbf{p}\) and \(\mathbf{q}\). \(R^2\) is often taught in
the context of OLS where \(y_i, f_i \in \mathbb{R}_1\). In that case,
\(d(y_i, f_i) = \sqrt{(y_i - f_i)^2}\); by extension
\(d(y_i, \bar{y}) = \sqrt{(y_i - \bar{y})^2}\). In the multidimensional
case where \(\mathbf{y}_i, \mathbf{f}_i \in \mathbb{R}_M; M > 1\), then
\(\bar{\mathbf{y}} \in \mathbb{R}_M\) represents the point at the center
of the data in \(\mathbb{R}_M\).\footnote{In the one-dimensional case,
  \(y_i , f_i \in \mathbb{R}_1\), \(SS_{resid.}\) can be considered the
  squared-Euclidean distance between the \(n\)-dimensional vectors \(y\)
  and \(f\). However, this relationship does not hold when
  \(\mathbf{y}_i , \mathbf{f}_i \in \mathbb{R}_M ; M > 1\).}

We can rewrite \(R^2\) using the relationships above. Note than now
\(\bar{\mathbf{y}}\) is now a vector with \(M\) entries. The \(j\)-th
entry of \(\bar{\mathbf{y}}\) is averaged across all \(N\) vectors. i.e
\(\bar{y}_j = \frac{1}{N} \sum_{i=1}^{N} y_{i,v}\). From there we have:

\begin{align}
    \bar{\mathbf{y}} &= \frac{1}{N} \sum_{i=1}^{N} \mathbf{y}_i \\ 
    SS_{tot.} &= \sum_{i=1}^N{d(\mathbf{y}_i, \bar{\mathbf{y}}})^2\\
    SS_{resid.} &= \sum_{i=1}^N{d(\mathbf{y}_i, \mathbf{f}_i)^2}\\
    \Rightarrow R^2 & \equiv 1 - \frac{SS_{resid.}}{SS_{tot.}}
\end{align}

Fig. 1 visualizes the geometric interpretation of \(R^2\) for outcomes
in \(\mathbb{R}_2\). The left image represents \(SS_{tot.}\): the red
dots are data points (\(\mathbf{y}_i\)); the black dot is the mean
(\(\bar{\mathbf{y}}\)); the line segments represent the Euclidean
distance from each \(\mathbf{y}_i\) to \(\bar{\mathbf{y}}\).
\(SS_{tot.}\) is obtained by squaring the length of each line segment
and then adding the squared segments together. The right image
represents \(SS_{resid.}\): the blue dots are the fitted values under
the model (\(\mathbf{f}_i\)); the line segments represent the Euclidean
distance from each \(\mathbf{f}_i\) to its corresponding
\(\mathbf{y}_i\). \(SS_{resid.}\) is obtained by squaring the length of
each line segment and then adding the squared segments together.

\begin{figure}
\centering
\includegraphics{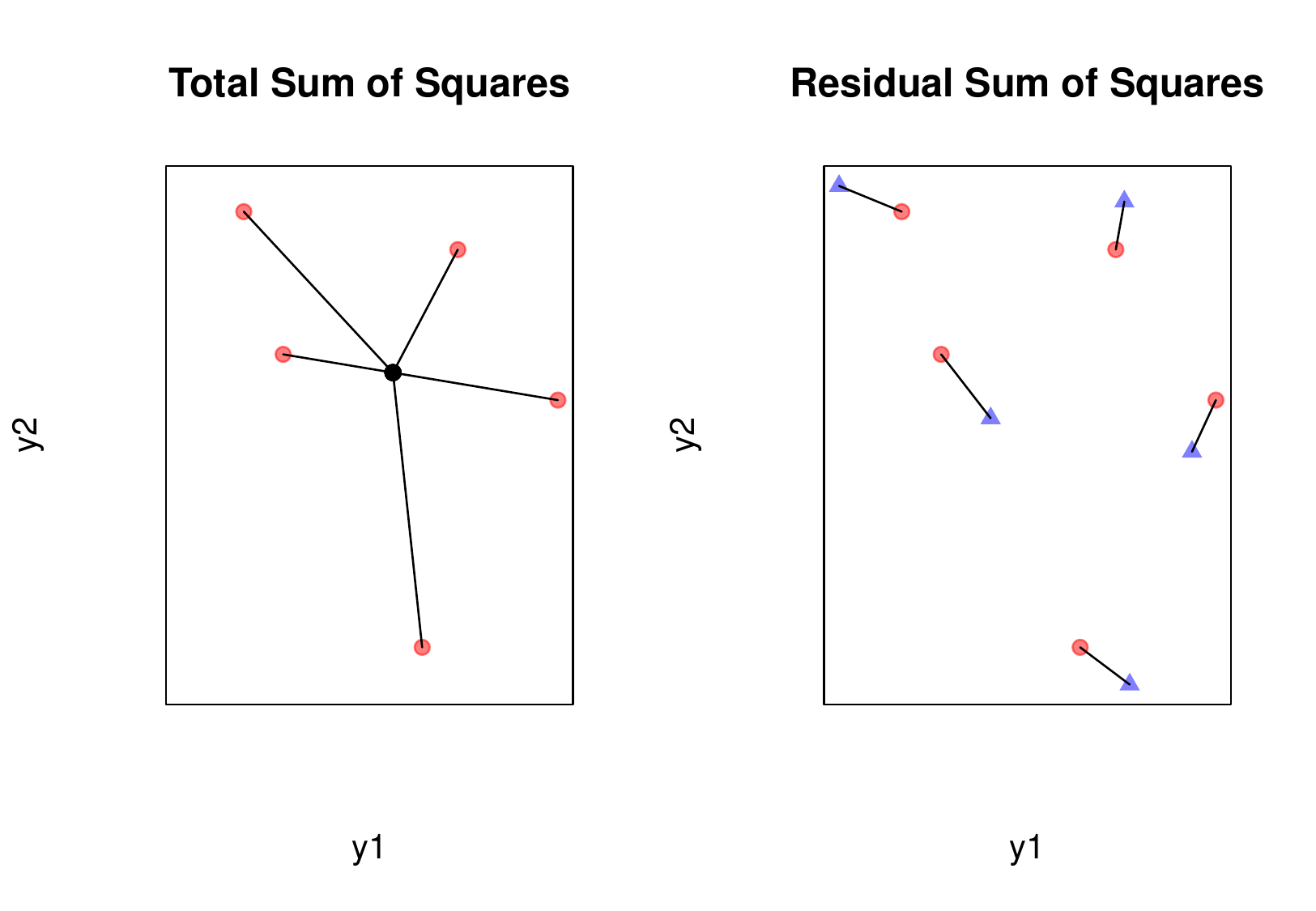}
\caption{Visualizing the geometric interpretation of R-squared:
corresponds to an R-squared of 0.87}
\end{figure}

The geometric interpretation of \(R^2\) is similar to the
``explained-variance'' interpretation. When \(SS_{resid.} = 0\), then
the model is a perfect fit for the data and \(R^2 = 1\). If
\(SS_{resid.} = SS_{tot.}\), then \(R^2 = 0\) and the model is no better
than just guessing \(\bar{y}\). When \(0 < SS_{resid} < SS_{tot}\), then
the model is a better fit for the data than a naive guess of
\(\bar{\mathbf{y}}\). In a non-linear or multi-dimensional model, it is
possible for \(SS_{resid.} > SS_{tot.}\). In this case, \(R^2\) is
negative, and guessing \(\bar{\mathbf{y}}\) is better than using the
model.

\hypertarget{extending-r2-to-topic-models}{%
\subsection{\texorpdfstring{Extending \(R^2\) to Topic
Models}{Extending R\^{}2 to Topic Models}}\label{extending-r2-to-topic-models}}

An \(R^2\) for topic models follows from the geometric interpretation of
\(R^2\). For a document, \(d\), the observed value, \(\mathbf{y}_d\), is
a vector of integers counting the number of times each token appears in
\(n_d\) draws. The document's fitted value under the model follows that
\(\mathbf{y}_d = \mathbf{w}_d\) represents the outcome of a multinomial
random variable. The fitted value is

\begin{align}
  \mathbf{f}_d = 
    \mathbb{E}(\mathbf{w}_d) = 
    n_d \odot \boldsymbol\theta_d \cdot \boldsymbol\Phi 
\end{align}

The center of the documents in the corpus, \(\bar{\mathbf{y}}\), is
obtained by averaging the token counts across all documents. From this
we obtain \(R^2\).

\begin{align}
    \bar{\mathbf{y}} &= \frac{1}{D}\sum_{d=1}^{D}\mathbf{y}_d\\
    SS_{tot.} &= \sum_{d=1}^D{d(\mathbf{y}_d, \bar{\mathbf{y}})^2}\\
    SS_{resid.} &= \sum_{d=1}^D{d(\mathbf{y}_d, \mathbf{f}_d)^2}\\
    R^2 &\equiv 1 - \frac{SS_{resid.}}{SS_{tot.} }
\end{align}

\hypertarget{pseudo-coefficients-of-variation}{%
\subsection{Pseudo Coefficients of
Variation}\label{pseudo-coefficients-of-variation}}

Several pseudo coefficients of variation have been developed for models
where the traditional \(R^2\) is inappropriate. Some of these, such as
Cox and Snell's \(R^2\) (Cox and Snell 1989) or McFadden's \(R^2\)
(McFadden, Tye, and Train 1977) may apply to topic models. As pointed
out in \emph{UCLA's Institute for Digital Research and Education},
(Bruin 2006)

\begin{quote}
These are `pseudo' R-squareds because they look like R-squared in the sense that they are on a similar scale, ranging from 0 to 1 (though some pseudo R-squareds never achieve 0 or 1) with higher values indicating better model fit, but they cannot be interpreted as one would interpret an OLS R-squared and different pseudo R-squareds can arrive at very different values. 
\end{quote}

The empirical section of this paper calculates an uncorrected McFadden's
\(R^2\) for topic models to compare to the standard (non-pseudo)
\(R^2\). McFadden's \(R^2\) is defined as

\begin{align}
    R^2_{Mc} & \equiv 1 - \frac{ ln( L_{full} ) }{ ln( L_{restricted} ) }
\end{align}

where \(L_{full}\) is the estimated likelihood of the data under the
model and \(L_{restricted}\) is the estimated likelihood of the data
free of the model. In the context of OLS, the restricted model is a
regression with only an intercept term. For other types of models (such
as topic models), care should be taken in selecting what ``free of the
model'' means.

For topic models, ``free of the model'' may mean that the words were
drawn from a simple multinomial distribution, whose parameter is
proportional to the relative frequencies of words in the corpus overall.
This is the specification used for the emperical analysis in this paper.

\hypertarget{empirical-evaluation-of-topic-model-r2}{%
\section{\texorpdfstring{Empirical Evaluation of Topic Model
\(R^2\)}{Empirical Evaluation of Topic Model R\^{}2}}\label{empirical-evaluation-of-topic-model-r2}}

This paper performs three analyses to empirically evaluate \(R^2\) for
topic models. Two analyses use Monte Carlo-simulated corpora and one
uses a corpus of grants awarded through the Department of Health and
Human Services. This latter corpus was obtained from the National
Institutes of Health \emph{NIH ExPORTER} database. (``Welcome to
Exporter,'' n.d.) The first analysis uses simulated corpora to observe
how the properties of training corpora influence \(R^2\). The second
analysis uses these same simulated corpora to compare \(R^2\) as
commonly-defined to McFadden's \(R^2\) in the context of topic modeling.
The final analysis compares the \(R^2\) values of various models
constructed on the NIH corpus.

\hypertarget{monte-carlo-simulated-corpora}{%
\subsection{Monte Carlo-Simulated
Corpora}\label{monte-carlo-simulated-corpora}}

It is possible to simulate corpora that share some key statistical
properties of human-generated language using the functional form of a
topic model. To do so, one must set \(\boldsymbol\beta\) such that its
entries are proportional to a power law. The result is a corpus whose
relative term frequencies follow Zipf's law of language (Zipf 1949). The
derivation of this result is included in the appendix.

This method generates a corpus of \(D\) documents, \(V\) tokens, and
\(K\) topics through the following stochastic process:

\begin{enumerate}
\def\labelenumi{\arabic{enumi}.}
\tightlist
\item
  Initialize
  \(\\ \boldsymbol\phi_k \sim \text{Dirichlet}(\boldsymbol\beta)\\\)
  \(\boldsymbol\theta_d \sim \text{Dirichlet}(\boldsymbol\alpha)\)
\item
  Then for each document draw \(\\ n_d \sim \text{Poisson}(\lambda)\)
\item
  Finally, for each document draw the following \(n_d\) times
  \(\\ z_{d,n} \sim \text{Multinomial}(1, \boldsymbol\theta_d)\\\)
  \(w_{d,n} \sim \text{Multinomial}(1, \boldsymbol\phi_{z_{d,n}})\)
\end{enumerate}

The words for document \(d\) are populated by sampling with replacement
from \(z_{d,n}\) and \(w_{dk,n}\) for \(n_d\) iterations. The parameters
\(V\), \(K\), and \(\lambda\) may be varied to adjust the corpus
properties for the number of tokens, topics, and average document length
respectively. Adjusting the shape and magnitude of \(\boldsymbol\alpha\)
and \(\boldsymbol\beta\) affect the concentration of topics within
documents and words within topics respectively.

\hypertarget{latent-dirichlet-allocation}{%
\subsection{Latent Dirichlet
Allocation}\label{latent-dirichlet-allocation}}

This paper uses Latent Dirichlet Allocation (LDA) and collapsed Gibbs
sampling to estimate topic models from actual data. LDA---possibly the
most popular topic model---is a Bayesian hierarchical model that places
Dirichlet priors on \(\boldsymbol\theta_d, \forall d\) and
\(\boldsymbol\phi_k, \forall k\) (Blei, Ng, and Jordan 2003).

A limitation with LDA is that the number of topics in the corpus must be
specified a priori where no prior knowledge often exists. Nevertheless,
LDA's empirical utility has been overwhelmingly demonstrated. For LDA
models fit in this paper, symmetric parameters are used for both prior
distributions. Each entry of \(\boldsymbol\alpha\) is 0.1; each entry of
\(\boldsymbol\beta\) is 0.01. Models are trained using Gibbs sampling
from the \emph{textmineR} package for the R language (Jones, n.d.).

\hypertarget{comparing-r2-to-mcfaddens-r2}{%
\subsection{\texorpdfstring{Comparing \(R^2\) to McFadden's
\(R^2\)}{Comparing R\^{}2 to McFadden's R\^{}2}}\label{comparing-r2-to-mcfaddens-r2}}

McFadden's pseudo \(R^2\) is calculated for all simulated corpora for
comparison to the standard \(R^2\). McFadden's \(R^2\) is a ratio of
likelihoods. Various methods exist for calculating likelihoods of topic
models (Buntine 2009). Most of these methods have a Bayesian perspective
and incorporate prior information. Not all topic models are Bayesian,
however. And in the case of simulated corpora, the exact data-generating
parameters are known a priori. As a result, likelihoods calculated for
this paper follow the simplest definition: they represent the
probability of observing the generated data, given the (known or
estimated) multinomial parameters of the model. The ``model-free''
likelihood assumes that each document is generated by drawing from a
single multinomial distribution. The parameters of this ``model-free''
distribution are proportional to the frequency of each token in the
data.

\hypertarget{empirical-properties-of-r2-for-topic-models}{%
\subsection{\texorpdfstring{Empirical Properties of \(R^2\) for Topic
Models}{Empirical Properties of R\^{}2 for Topic Models}}\label{empirical-properties-of-r2-for-topic-models}}

The \(R^2\) for topic models has the following empirical properties: It
is bound between \(-\infty\) and 1. \(R^2\) is invariant to the
\textit{true} number of topics in the corpus. \(R^2\) increases with the
\textit{estimated} number of topics using LDA; this indicates that there
might be a risk of overfit from a model with too many topics (discussed
in more detail in the next section). \(R^2\) decreases as the vocabulary
size of the corpus increases. \(R^2\) increases as the average document
length increases. (See Fig. 2, Fig. 3, Fig. 4, and Fig. 5.)

\begin{figure}
\centering
\includegraphics{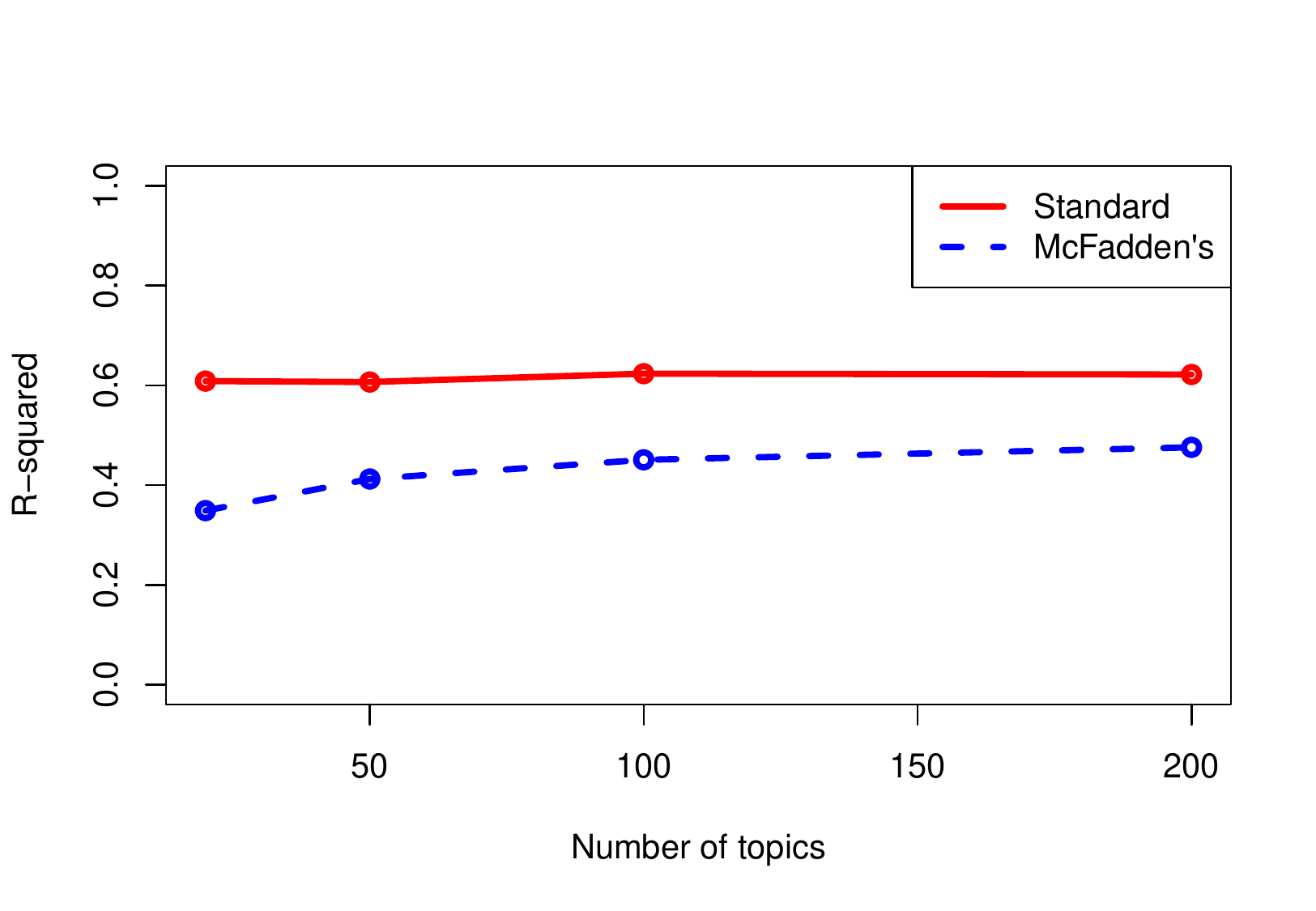}
\caption{Varying the number of topics on a simulated corpus}
\end{figure}

\begin{figure}
\centering
\includegraphics{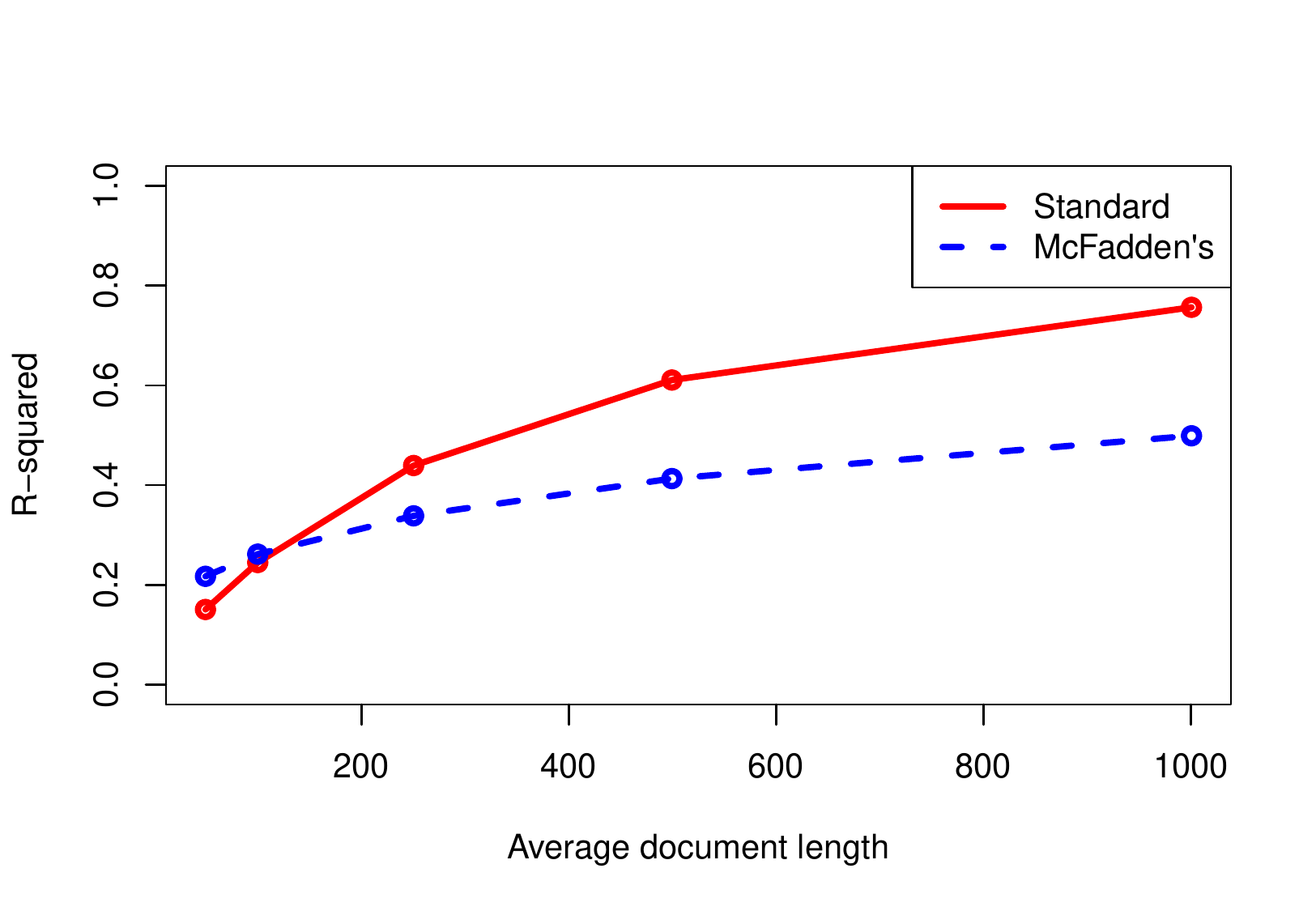}
\caption{Varying average document length on a simulated corpus}
\end{figure}

\begin{figure}
\centering
\includegraphics{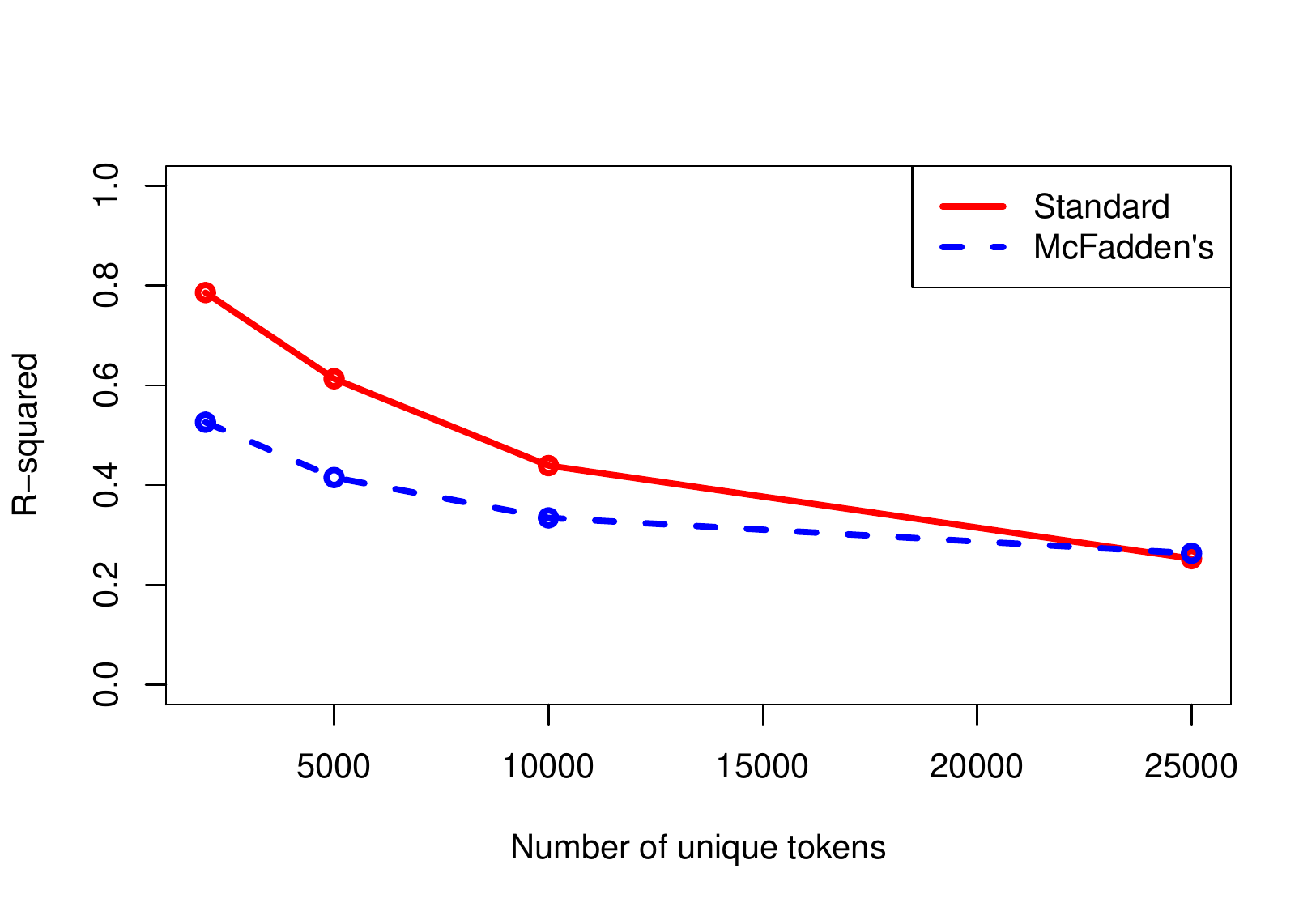}
\caption{Varying vocabulary size on a simulated corpus}
\end{figure}

\begin{figure}
\centering
\includegraphics{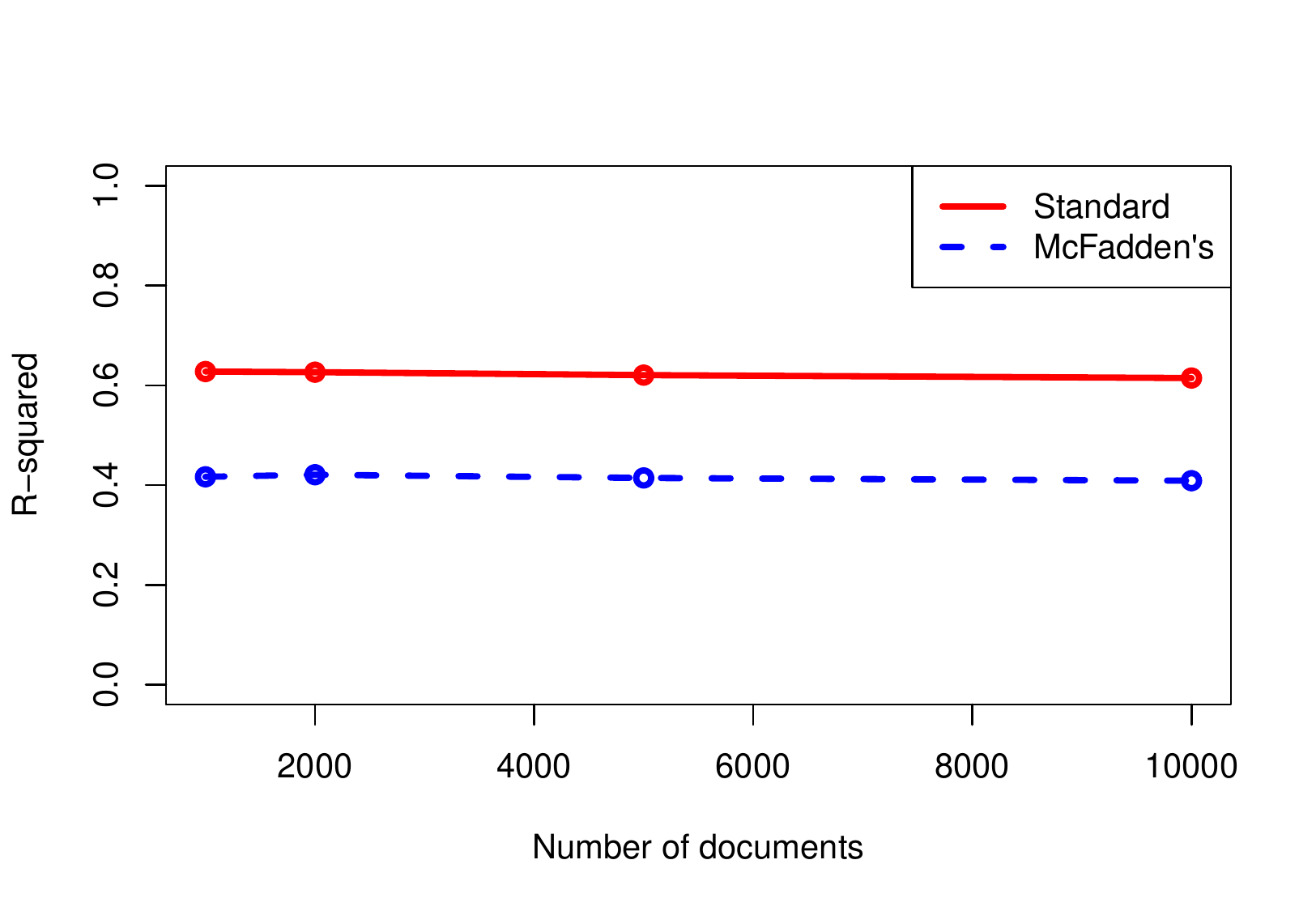}
\caption{Varying number of documents on a simulated corpus}
\end{figure}

Most of these empirical properties were obtained by using simulated
corpora, as described earlier in the paper, with one exception. The
default parameter settings for Monte Carlo simulation are \(K = 50\)
topics, \(D = 2{,}000\) documents, \(V = 5{,}000\) tokens, and
\(\lambda = 500\) for the average document length, which is distributed
\(\text{Poisson}(\lambda)\). Each parameter was varied, holding other
parameters constant. In each case, Monte Carlo simulation generates a
document term matrix for the corpus while the parameters for generating
the documents are known. \(R^2\) is calculated for each simulated
corpus, using the population parameters. These \(R^2\) metrics represent
a best-case scenario, avoiding misspecification and other pathologies
present with topic model estimation algorithms.

McFadden's pseudo \(R^2\) is also calculated for these simulated data.
McFadden's \(R^2\) is generally lower than the standard \(R^2\).
McFadden's \(R^2\) is subject to the common problem of many pseudo
\(R^2\) measures; its true upper bound is less than one. This makes
sense. If McFadden's \(R^2\) were to equal 1, then the likelihood of the
data would be 1 which is impossible. It is never the case that a
likelihood will equal 1. Given the scale of linguistic data, the
likelihood will always be significantly less than 1. Therefore, a scale
correction measure is needed, making McFadden's \(R^2\) more
complicated.

McFadden's \(R^2\) increases slightly with the number of true topics;
this is problematic. When the scale of an \(R^2\) varies with
\textit{known} properties of the data, such as the number of documents,
vocabulary size, average document length, etc. scale correction measures
are possible. However, when the metric varies with an \textit{unknown}
property, such as the number of latent topics, then a scale correction
is not possible. For this reason alone, McFadden's \(R^2\) is
undesirable. Other properties of McFadden's \(R^2\) are consistent with
the properties of the standard \(R^2\)

These properties of \(R^2\) suggest a change in focus for the topic
modeling research community. Document length is an important factor in
model fit whereas the number of documents is not. When choosing between
fitting a topic model on short abstracts, full-page documents, or
multi-page papers, it appears that more text is better. Second, since
model fit is invariant for corpora over 1,000 documents (our lower bound
for simulation), taking a sample from a large corpus may yield
reasonable estimates of the larger corpus's parameters.\footnote{Estimating
  appropriate sample sizes and strategies based on properties of the
  text and desired analyses is a fruitful area of future research.} The
topic modeling community has heretofore focused on scalability to large
corpora of hundreds-of-thousands to millions of documents. There has
been less focus on document length.

\hypertarget{comparison-of-simulated-data-with-the-nih-corpus}{%
\subsection{Comparison of Simulated Data with the NIH
Corpus}\label{comparison-of-simulated-data-with-the-nih-corpus}}

\(R^2\) increases with the \textit{estimated} number of topics using
LDA. This indicates a risk of model overfit with respect to the number
of estimated topics. To evaluate the effect of \(R^2\) on the number of
estimated topics, LDA models were fit to two corpora. The first corpus
is simulated, using the parameter defaults: \(K = 50\), \(D = 2{,}000\),
\(V = 5{,}000\), and \(\lambda = 500\). The second corpus is on the
abstracts of 1,000 randomly-sampled research grants for fiscal year 2014
in the National Institutes of Health's ExPORTER database. In both cases,
LDA models are fit to the data estimating a range of \(K\). For each
model, \(R^2\) and the log likelihood are calculated. The known
parameters for this NIH corpus are \(D = 1{,}000\), \(V = 8,751\), and
the median document length is 222 words. This vocabulary has standard
stop words removed,\footnote{175 English stop words from the snowball
  stemmer project.
  http://snowball.tartarus.org/algorithms/english/stop.txt} is tokenized
to consider only unigrams, and excludes tokens that appear in fewer than
2 documents.

The same likelihood calculation is used here, as described above. This
likelihood calculation excludes prior information typically used when
calculating the likelihood of an LDA model. This is perhaps not the
optimal method for calculating likelihoods when using a Bayesian model.
However, this prior information is excluded here for two reasons. First,
it is consistent with the method used for calculating McFadden's \(R^2\)
earlier in the paper. Second, this log likelihood can be calculated
exactly. LDA's likelihood is contained within an intractable integral.
Various approximations have been developed (Buntine 2009). However,
there is some risk that a comparison of an approximated likelihood to
\(R^2\) may be biased by the approximation method.

\begin{figure}
\centering
\includegraphics{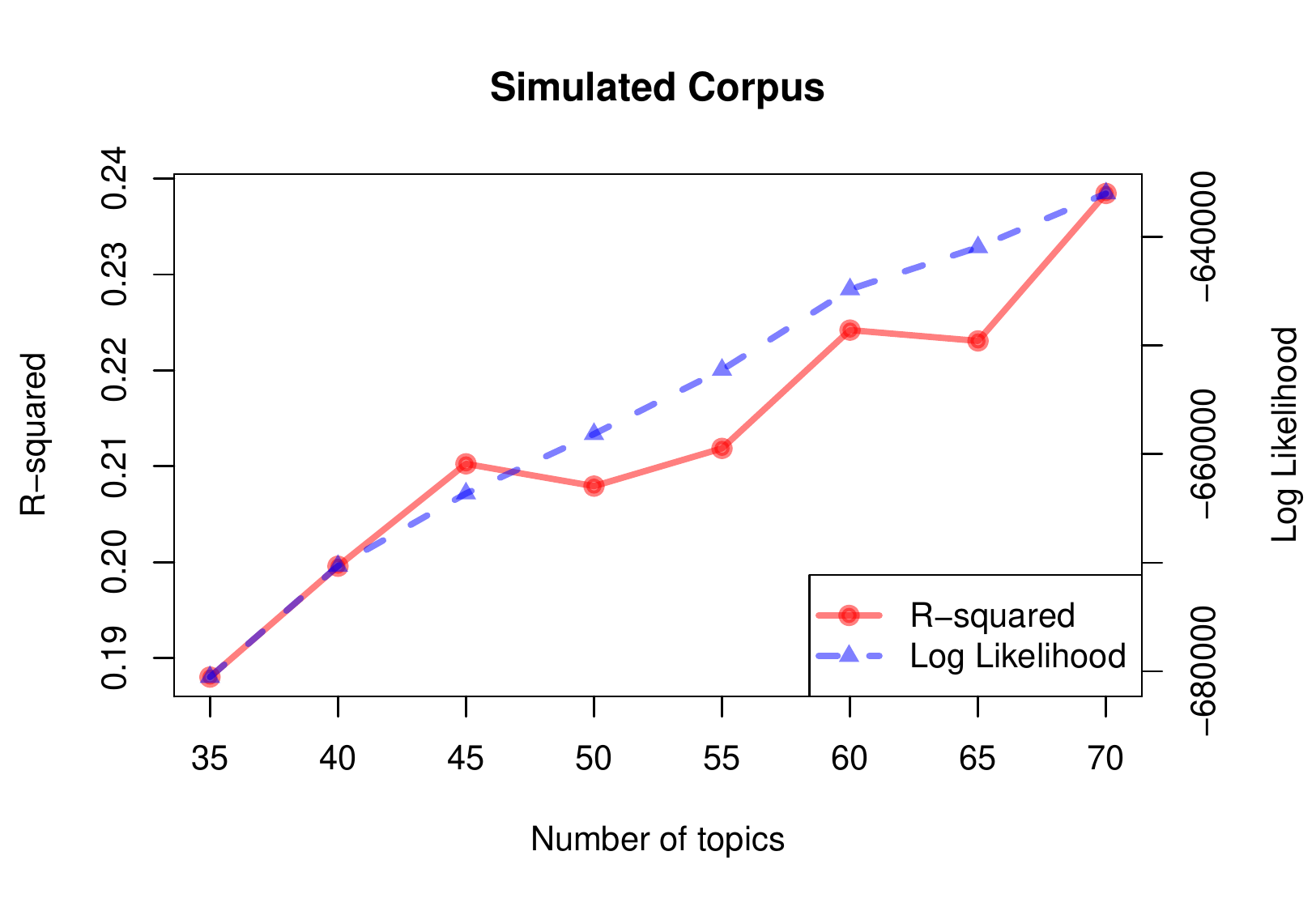}
\caption{Comparison of R2 and log likelihood for LDA models fit on a
simulated corpus}
\end{figure}

\begin{figure}
\centering
\includegraphics{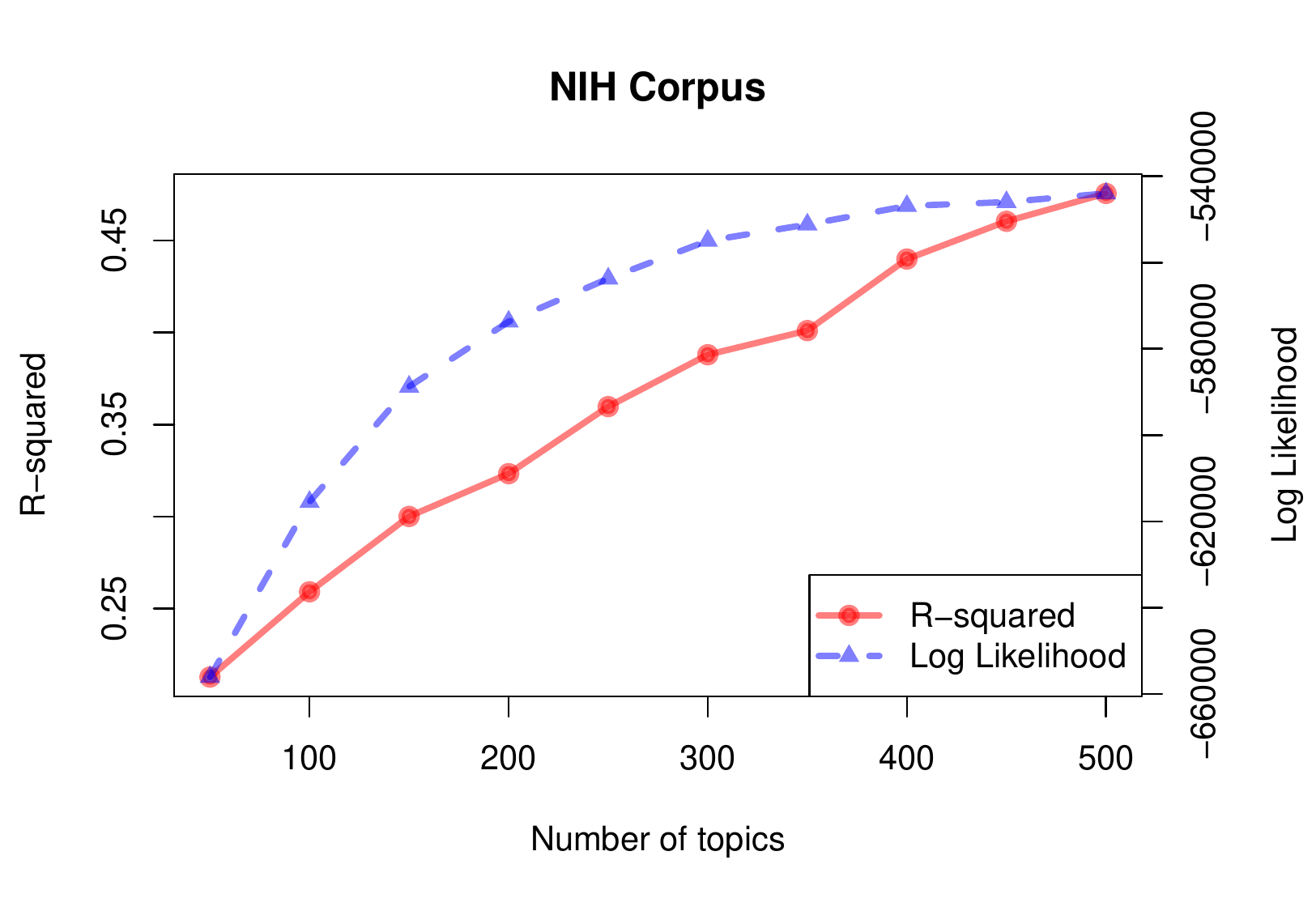}
\caption{Comparison of R2 and log likelihood for LDA models fit on the
NIH corpus}
\end{figure}

Fig. 6 and Fig. 7 depict \(R^2\) and the log likelihood over a range of
estimated \(K\) for both corpora. Fig. 6 corresponds to the simulated
corpus. Fig. 7 corresponds to the NIH corpus. From both images, we see
that \(R^2\) and the log likelihood both increase with the number of
estimated topics. The scale of \(R^2\) is comparable between the
simulated corpus and the NIH corpus, in spite of their differences in
size. In the case of the simulated corpus, we know that the true number
of topics is 50. However, this is not clear from observing \(R^2\) or
the log likelihood. It does not appear that \(R^2\) or this ``raw''
likelihood can help find the true number of topics.

Chang et al. (2009) observe that there may be a trade-off between model
fit and human interpretation. Specifically, they find that humans can
more easily interpret models with fewer topics. This may be true. As
discussed in an earlier section, non-\(R^2\) goodness of fit measures
are not readily comparable across corpora and models. Because \(R^2\) is
comparable, it is now possible to quantify how much goodness of fit is
lost when K is lowered. This does not address any issues arising from a
pathological misspecification of the model, an ``incorrect'' K, for
example. In the case of the NIH corpus, \(R^2\) goes from 0.27 to 0.2
when the number of estimated topics is lowered from 200 to 100.

\hypertarget{conclusion}{%
\section{Conclusion}\label{conclusion}}

\(R^2\) has many advantages over standard goodness of fit measures
commonly used in topic modeling. Current goodness of fit measures are
difficult to interpret, compare across corpora, and explain to lay
audiences. \(R^2\) does not have any of these issues. Its scale is
effectively bounded between 0 and 1, as negative values (though
possible) are rare and indicate extreme model misspecification. \(R^2\)
may be used to compare models of different corpora, if necessary.
Scientifically-literate lay audiences are almost uniformly familiar with
\(R^2\) in the context of linear regression; the topic model \(R^2\) has
a similar interpretation, making it an intuitive measure.

The standard (geometric) interpretation of \(R^2\) is preferred to
McFadden's pseudo \(R^2\). The effective upper bound for McFadden's
\(R^2\) is considerably smaller than 1. A scale correction measure is
needed. Also, it is debatable which likelihood calculation(s) are most
appropriate. These issues make McFadden's \(R^2\) complicated and
subjective. However, a primary motivation for deriving a topic model
\(R^2\) is to remove the complications that currently hinder evaluating
and communicating the fidelity with which topic models represent
observed text. Most problematically, McFadden's \(R^2\) varies with the
number of \textit{true} topics in the data. It is therefore unreliable
in practice where the true number of topics in unknown.

One result of this research indicates that further study is needed on
the relationship between goodness of fit and document length, the number
of documents in a corpus, and vocabulary size. Results reported in this
paper demonstrate that document length is a considerable factor in model
fit, whereas the number of documents (above 1,000) is not. If robust,
this result indicates that the topic modeling community may need to
change focus away from scaling estimation algorithms for large corpora.
Instead, more effort should be put towards obtaining high-quality data.
Also, studying the relationship between these parameters, along with the
number of topics, may facilitate the development of an adjusted \(R^2\),
guarding against model overfit.

Lack of consistent evaluation metrics has limited the use of topic
models as a mature statistical method. The development of an \(R^2\) for
topic modeling is no silver bullet. However it represents a step towards
establishing consistency and rigor in topic modeling. This paper
proposes reporting \(R^2\) as a standard metric alongside topic models,
as is typically done with OLS.

\newpage

\hypertarget{appendix}{%
\section{Appendix}\label{appendix}}

Below derives the expected term frequency of a corpus whose terms are
generated by the stochastic process modeled by Latent Dirichlet
Allocation. This process is

\begin{enumerate}
\def\labelenumi{\arabic{enumi}.}
\tightlist
\item
  Initialize
  \(\\ \boldsymbol\phi_k \sim \text{Dirichlet}(\boldsymbol\beta)\\\)
  \(\boldsymbol\theta_d \sim \text{Dirichlet}(\boldsymbol\alpha)\)
\item
  Then for each document draw \(\\ n_d \sim \text{Poisson}(\lambda)\)
\item
  Finally, for each document draw the following \(n_d\) times
  \(\\ z_{d,n} \sim \text{Multinomial}(1, \boldsymbol\theta_d)\\\)
  \(w_{d,n} \sim \text{Multinomial}(1, \boldsymbol\phi_{z_{d,n}})\)
\end{enumerate}

The expected term frequencies of a corpus generated with the above
process are proportional to \(\boldsymbol\beta\)---the parameter for the
Dirichlet prior for terms over topics. This implies that for a simulated
corpus to follow Zipf's law, then \(\boldsymbol\beta\) must be
proportional to a power law.\footnote{There may be other implications as
  well. The author plans to explore these in future research.}

Let's start by carrying the expected value through both sides of
equation (11), above, using the law of total expectation.

\begin{align*}
  \mathbb{E}(\mathbf{w}_d) 
    &= \mathbb{E}\left(n_d \odot 
      \boldsymbol\theta_d \cdot \boldsymbol\Phi\right)\\
    &= \mathbb{E}\left(n_d
      \begin{pmatrix}
        \sum_{k = 1}^K \theta_{d,k} \cdot \phi_{k,1}\\
        \sum_{k = 1}^K \theta_{d,k} \cdot \phi_{k,2}\\
        ...\\
        \sum_{k = 1}^K \theta_{d,k} \cdot \phi_{k,V}\\
      \end{pmatrix}
    \right)\\
    &= \mathbb{E}(n_d)
      \begin{pmatrix}
        \mathbb{E}(\sum_{k = 1}^K \theta_{d,k} \cdot \phi_{k,1})\\
        \mathbb{E}(\sum_{k = 1}^K \theta_{d,k} \cdot \phi_{k,2})\\
        ...\\
        \mathbb{E}(\sum_{k = 1}^K \theta_{d,k} \cdot \phi_{k,V})\\
      \end{pmatrix}\\
    &= \mathbb{E}(n_d)
      \begin{pmatrix}
        \sum_{k = 1}^K \mathbb{E}(\theta_{d,k}) \mathbb{E}(\phi_{k,1})\\
        \sum_{k = 1}^K \mathbb{E}(\theta_{d,k})\mathbb{E}(\phi_{k,2})\\
        ...\\
        \sum_{k = 1}^K \mathbb{E}(\theta_{d,k}) \mathbb{E}(\phi_{k,V})\\
      \end{pmatrix}
\end{align*}

The last step, above, is due to indpendence of \(\boldsymbol\theta_d\)
and \(\boldsymbol\phi_k\) \(\forall d,k\).

Before carrying on, let's note two more relationships:

\begin{enumerate}
\def\labelenumi{\arabic{enumi}.}
\tightlist
\item
  \(\boldsymbol\phi_k \sim \text{i.i.d. Dirichlet}(\boldsymbol\beta)\)
  means that
  \(\mathbb{E}(\boldsymbol\phi_i) = \mathbb{E}(\boldsymbol\phi_j)\)
  \(\forall i,j \in \{1,2,..,K\}\)
\item
  The expected value of a Dirchlet random
  variable---\(\mathbf{X}\)---with parameter \(\boldsymbol\delta\) is
  \(\mathbb{E}(\mathbf{X}) = \frac{1}{\sum_{m=1}^M\delta_m}\cdot\boldsymbol\delta\)
\end{enumerate}

From number 1., above, we can pull \(\mathbb{E}(\phi_{.,k})\) outside of
the summation. And we can carry through the expected values using
definition in 2., above.

\begin{align*}
  \mathbb{E}(\mathbf{w}_d) 
    &= \mathbb{E}(n_d)
      \begin{pmatrix}
        \mathbb{E}(\phi_{k,1}) \sum_{k = 1}^K \mathbb{E}(\theta_{d,k}) \\
        \mathbb{E}(\phi_{k,2})\sum_{k = 1}^K \mathbb{E}(\theta_{d,k})\\
        ...\\
        \mathbb{E}(\phi_{k,V})\sum_{k = 1}^K \mathbb{E}(\theta_{d,k}) \\
      \end{pmatrix}\\
    &= n_d
      \begin{pmatrix}
        \frac{\beta_1}{\sum_{v=1}^V\beta_v} 
          \sum_{k = 1}^K \frac{\alpha_k}{\sum_{k = 1}^K \alpha_k} \\
        \frac{\beta_2}{\sum_{v=1}^V\beta_v} 
          \sum_{k = 1}^K \frac{\alpha_k}{\sum_{k = 1}^K \alpha_k} \\
        ...\\
        \frac{\beta_V}{\sum_{v=1}^V\beta_v} 
          \sum_{k = 1}^K \frac{\alpha_k}{\sum_{k = 1}^K \alpha_k} \\
      \end{pmatrix}\\
    &= n_d
      \begin{pmatrix}
        \frac{\beta_1}{\sum_{v=1}^V\beta_v} 
          \frac{\sum_{k = 1}^K \alpha_k}{\sum_{k = 1}^K \alpha_k} \\
        \frac{\beta_2}{\sum_{v=1}^V\beta_v} 
          \frac{\sum_{k = 1}^K \alpha_k}{\sum_{k = 1}^K \alpha_k} \\
        ...\\
        \frac{\beta_V}{\sum_{v=1}^V\beta_v} 
          \frac{\sum_{k = 1}^K \alpha_k}{\sum_{k = 1}^K \alpha_k} \\
      \end{pmatrix}\\
    &= n_d
      \begin{pmatrix}
        \frac{\beta_1}{\sum_{v=1}^V\beta_v} \cdot 1 \\
        \frac{\beta_2}{\sum_{v=1}^V\beta_v} \cdot 1  \\
        ...\\
        \frac{\beta_V}{\sum_{v=1}^V\beta_v} \cdot 1  \\
      \end{pmatrix}\\
   &= \frac{n_d}{\sum_{v=1}^V\beta_v}\boldsymbol\beta \\
   &\propto \boldsymbol\beta
\end{align*}

The end result is that the expected term frequency of a single document
is proportional to \(\boldsymbol\beta\)---the Dirichlet parameter for
terms over topics.

The term frequency for the whole corpus is the sum of the term
frequencies for each document. Specifically

\begin{align*}
  \mathbf{w} = \sum_{d=1}^D\mathbf{w}_d
\end{align*}

The expected value under the model, then, can be carried through.

\begin{align*}
  \mathbb{E}(\mathbf{w})
    &= \mathbb{E}\left(\sum_{d=1}^D\mathbf{w}_d\right) \\
    &= \sum_{d=1}^D\mathbb{E}(\mathbf{w}_d)\\
    &= \sum_{d=1}^D \frac{n_d}{\sum_{v=1}^V\beta_v}\boldsymbol\beta\\
    &= \frac{\sum_{d=1}^D n_d}{\sum_{v=1}^V\beta_v}\boldsymbol\beta\\
    &\propto \boldsymbol\beta
\end{align*}

\newpage

\hypertarget{references}{%
\section*{References}\label{references}}
\addcontentsline{toc}{section}{References}

\hypertarget{refs}{}
\leavevmode\hypertarget{ref-asuncion2009smoothing}{}%
Asuncion, Arthur, Max Welling, Padhraic Smyth, and Yee Whye Teh. 2009.
``On Smoothing and Inference for Topic Models.'' In \emph{Proceedings of
the Twenty-Fifth Conference on Uncertainty in Artificial Intelligence},
27--34. AUAI Press.

\leavevmode\hypertarget{ref-blei2006dynamic}{}%
Blei, David M, and John D Lafferty. 2006. ``Dynamic Topic Models.'' In
\emph{Proceedings of the 23rd International Conference on Machine
Learning}, 113--20. ACM.

\leavevmode\hypertarget{ref-blei2003latent}{}%
Blei, David M, Andrew Y Ng, and Michael I Jordan. 2003. ``Latent
Dirichlet Allocation.'' \emph{Journal of Machine Learning Research} 3
(Jan): 993--1022.

\leavevmode\hypertarget{ref-bruin2006faq}{}%
Bruin, J. 2006. ``FAQ. What Are Pseudo-R-Squareds.'' UCLA: Academic
Technology Services, Statistical Consulting Group. Retrieved~\ldots{}.

\leavevmode\hypertarget{ref-buntine2009estimating}{}%
Buntine, Wray. 2009. ``Estimating Likelihoods for Topic Models.'' In
\emph{Asian Conference on Machine Learning}, 51--64. Springer.

\leavevmode\hypertarget{ref-chang2009reading}{}%
Chang, Jonathan, Sean Gerrish, Chong Wang, Jordan L Boyd-Graber, and
David M Blei. 2009. ``Reading Tea Leaves: How Humans Interpret Topic
Models.'' In \emph{Advances in Neural Information Processing Systems},
288--96.

\leavevmode\hypertarget{ref-cox1989analysis}{}%
Cox, DR, and EJ Snell. 1989. \emph{Analysis of Binary Data}. Vol. 32.
CRC Press.

\leavevmode\hypertarget{ref-girolami2003equivalence}{}%
Girolami, Mark, and Ata Kabán. 2003. ``On an Equivalence Between Plsi
and Lda.'' In \emph{SIGIR}, 3:433--34.

\leavevmode\hypertarget{ref-griffiths2004finding}{}%
Griffiths, Thomas L, and Mark Steyvers. 2004. ``Finding Scientific
Topics.'' \emph{Proceedings of the National Academy of Sciences} 101
(suppl 1). National Acad Sciences: 5228--35.

\leavevmode\hypertarget{ref-textminer}{}%
Jones, Tommy. n.d. ``TextmineR.'' \url{https://www.rtextminer.com}.

\leavevmode\hypertarget{ref-lau2014machine}{}%
Lau, Jey Han, David Newman, and Timothy Baldwin. 2014. ``Machine Reading
Tea Leaves: Automatically Evaluating Topic Coherence and Topic Model
Quality.'' In \emph{Proceedings of the 14th Conference of the European
Chapter of the Association for Computational Linguistics}, 530--39.

\leavevmode\hypertarget{ref-mcfadden1977application}{}%
McFadden, Daniel, William B Tye, and Kenneth Train. 1977. \emph{An
Application of Diagnostic Tests for the Independence from Irrelevant
Alternatives Property of the Multinomial Logit Model}. Institute of
Transportation Studies, University of California.

\leavevmode\hypertarget{ref-neter1996applied}{}%
Neter, John, Michael H Kutner, Christopher J Nachtsheim, and William
Wasserman. 1996. \emph{Applied Linear Statistical Models}. Vol. 4. Irwin
Chicago.

\leavevmode\hypertarget{ref-nguyen2014sometimes}{}%
Nguyen, Viet-An, Jordan Boyd-Graber, and Philip Resnik. 2014.
``Sometimes Average Is Best: The Importance of Averaging for Prediction
Using Mcmc Inference in Topic Modeling.'' In \emph{Proceedings of the
2014 Conference on Empirical Methods in Natural Language Processing},
1752--7.

\leavevmode\hypertarget{ref-roberts2014structural}{}%
Roberts, Margaret E, Brandon M Stewart, Dustin Tingley, Christopher
Lucas, Jetson Leder-Luis, Shana Kushner Gadarian, Bethany Albertson, and
David G Rand. 2014. ``Structural Topic Models for Open-Ended Survey
Responses.'' \emph{American Journal of Political Science} 58 (4). Wiley
Online Library: 1064--82.

\leavevmode\hypertarget{ref-rosner2014evaluating}{}%
Rosner, Frank, Alexander Hinneburg, Michael Röder, Martin Nettling, and
Andreas Both. 2014. ``Evaluating Topic Coherence Measures.'' \emph{arXiv
Preprint arXiv:1403.6397}.

\leavevmode\hypertarget{ref-wallach2009rethinking}{}%
Wallach, Hanna M, David M Mimno, and Andrew McCallum. 2009. ``Rethinking
Lda: Why Priors Matter.'' In \emph{Advances in Neural Information
Processing Systems}, 1973--81.

\leavevmode\hypertarget{ref-nih}{}%
``Welcome to Exporter.'' n.d. \url{https://exporter.nih.gov/}.

\leavevmode\hypertarget{ref-zipf1949human}{}%
Zipf, George Kingsley. 1949. ``Human Behavior and the Principle of Least
Effort.'' Addison-Wesley Press.

\end{document}